\documentclass[prd,aps,floats,preprint,preprintnumbers,amssymb,nofootinbib]{revtex4}

\usepackage{graphicx}
\usepackage{enumitem}
\usepackage{latexsym}
\usepackage{amsfonts}
\usepackage{amssymb}
\usepackage[dvipsnames]{xcolor}
\usepackage{CJK}
\usepackage[export]{adjustbox}
\usepackage{amsmath}
\usepackage{slashed}
\usepackage{dcolumn}
\usepackage{verbatim}
\usepackage{float}
\usepackage{multirow}
\usepackage{soul}
\usepackage[normalem]{ulem}
\usepackage{ulem}
\usepackage{hyperref}
\usepackage{tikz}
\usepackage{adjustbox}
\usepackage[compat=1.1.0]{tikz-feynman}
\usepackage{natbib}
\usepackage{epsfig}

\newcommand{\be}{\begin{equation}}
\newcommand{\ee}{\end{equation}}
\newcommand{\bea}{\begin{eqnarray}}
\newcommand{\eea}{\end{eqnarray}}

\newcommand{\mpl}{M_{\rm Pl}}

\newcommand{\dm}{{\rm dm}}
\newcommand{\ede}{{\rm ede}}

\begin{document}


\title{Coupled Early Dark Energy}
\author{Mark Trodden\footnote{trodden@physics.upenn.edu}}

\affiliation{
Center for Particle Cosmology, Department of Physics and Astronomy, University of Pennsylvania,
Philadelphia PA 19104, USA}

\date{\today}

\begin{abstract}
Early dark energy has emerged as one of the more promising approaches to address the Hubble tension - the statistically significant disparity between measurements of the Hubble constant made using data from different epochs in cosmic history. However, the idea is not without its own set of challenges, both from the data, in the effects it has on other measurements, such as the large-scale structure tension, and from theoretical concerns such as technical naturalness and the introduction of a new coincidence problem in cosmology. In this brief note, delivered as an invited plenary lecture at the {\it 15th Frontiers of Fundamental Physics conference}, I discuss how some of the fine-tuning problems of early dark energy can be ameliorated by using couplings to other fields already present in cosmology, and for which the epoch of matter-radiation equality is already a special one. The resulting models - neutrino assisted early dark energy, and chameleon early dark energy - provide testable, theoretically robust implementations of this general idea. I will discuss the formulation and the cosmology of such approaches, including some constraints arising from both observational and theoretical considerations.
\end{abstract}

\maketitle

\section{Introduction}
An issue of increasing interest in recent years is the disagreement between measurements of the Hubble constant made using two different methods - one (the lower estimate) made using observations of the cosmic microwave background (CMB)~\cite{Aghanim:2018eyx} and another (the higher estimate) made using both local distance indicators~\cite{Riess:2021jrx, Freedman:2020dne} and time-delays from strong lensing~\cite{Wong:2019kwg,Birrer:2020tax}. This issue is known as the {\it Hubble tension} \cite{Verde:2019ivm}, and has shown itself to be robust as new and more precise datasets have arrived~\cite{Fitzpatrick:2000hh,Benedict:2006cp,Humphreys:2013eja,Efstathiou:2013via,Rigault:2014kaa,Becker:2015nya,Shanks:2018rka,Riess:2018kzi,Kenworthy:2019qwq,Spergel:2013rxa}. 

Naturally, theorists are intrigued by this discrepancy since it holds the possibility of being a first hint of new physics beyond the standard model of cosmology. A particularly interesting proposal is that of early dark energy (EDE)~\cite{Karwal:2016vyq,Poulin:2018cxd}, in which a scalar field makes an important contribution to the cosmic energy budget around the time of matter-radiation equality, while being essentially irrelevant before and after this epoch. This raises the inferred value of the Hubble constant as measured through the CMB by reducing the sound horizon for acoustic waves, and hence the angular diameter distance to the surface of last scattering~\cite{Bernal:2016gxb,Evslin:2017qdn,Aylor:2018drw,Knox:2019rjx}. 

In this presentation, delivered as an invited plenary lecture at the {\it 15th Frontiers of Fundamental Physics conference}, I discussed two different theoretical approaches to addressing both the coincidence problem of EDE models (the necessity of arbitrarily choosing the mass of the scalar in order to ensure that it begins to roll close to matter-radiation equality), and the problems of naturalness, technical and otherwise, from which typical EDE models suffer. In particular, we note that the more successful uncoupled scalar-field EDE models have potentials such as $V(\phi) \sim (1 - \cos \phi)^n$ for integer $n$ \cite{Poulin:2018dzj,Smith:2019ihp}, which are difficult to theoretically motivate. Couplings to other fields can significantly ameliorate this problem.

The first of these approaches~\cite{Sakstein:2019fmf,CarrilloGonzalez:2020oac} exploits the coincidence between the upper limit on the sum of the neutrino masses ($\le 0.5$eV~\cite{Aghanim:2018eyx,Aker:2019uuj}) and the energy scale of matter-radiation equality. This {\it neutrino-assisted early dark energy} model couples the EDE field to neutrinos, resulting in an energy injection when neutrinos transition from being relativistic to non-relativistic, after which the EDE field then rolls down its potential and addresses the Hubble tension in much the same way as in regular EDE models.

The second approach~\cite{Karwal:2021vpk} instead exploits the fact that dark matter becomes the dominant component of the Universe close to matter-radiation equality. The resulting {\it Chameleon Early Dark Energy} models involve conformally coupling the EDE field to dark matter, in order to naturally ensure that the relevant dynamics are triggered at the correct epoch. 
The introduction of this coupling address not only the fine-tuning of the EDE injection time, but also various other criticisms of EDE models. 

I will review the essential features of these models and describe how the basic physics works using examples of various levels of complexity. I will then describe the background cosmological evolution, and present some rudimentary analyses of how effective the mechanisms are at addressing the Hubble tension, with a more complete analysis being beyond the scope of this talk. I
will particularly focus on the theoretical challenges that arise from treating these approaches as serious effective field theories, considering quantum corrections and naturalness issues. In places I have reused material from the two papers on which this talk was based. Furthermore, I have not attempted to be comprehensive in my references, instead listing only those that were relevant to the actual talk I delivered, leaving detailed referencing to the actual papers on which the talk is based.

\section{Neutrino-Assisted EDE}

EDE scenarios~\cite{Kamionkowski:2014zda} add a new scalar-field component to the Universe which is insignificantly subdominant at all times except for a redshift-localized contribution close to matter-radiation equality. 
From a particle-physics perspective, a serious drawback of such a field is the excessive fine-tuning of its parameters, including exceptionally small masses, required to achieve kination at the right moment in cosmic history. 
A natural question therefore is to consider whether EDE might be coupled to other components, such that its dynamics are triggered by some other, pre-existing physics concurrent with the epoch of equality. 

One way to do this~\cite{Sakstein:2019fmf,CarrilloGonzalez:2020oac} is through {\it neutrino-assisted early dark energy}, which makes use of the coincidence between the upper limit on the sum of the neutrino masses ($\leq 0.5$eV~\cite{Aghanim:2018eyx,Aker:2019uuj}) and the energy scale of matter-radiation equality (the neutrino temperature at $z=3000$ is $0.51$ eV.). The broad idea is that if neutrinos are conformally coupled to the EDE field, then this might allow us the freedom to consider much heavier EDE fields. This is because rather than having to use Hubble friction to freeze EDE fields on their potentials at equality, we can instead arrange for them to receive an energy injection, or ``kick", at this time, as neutrinos transition from being relativistic to non-relativistic. Once it has been displaced by the kick, the EDE field might behave in much the same way as in regular EDE models, rolling down its potential to its minimum. 

\subsection{The Basic Idea of $\nu$-assisted EDE}
The simplest incarnation of this model~\cite{Sakstein:2019fmf} consists of a single neutrino species, without any of the gauge symmetries of the standard model, allowing for a regular mass term, with action
\begin{equation}
\label{eq:act}
S=\int d^4 x\sqrt{-g}\left[\frac{\mpl^2}{2} R(g)-\frac12\nabla_\mu\phi\nabla^\mu\phi-V(\phi)\right]+S_\nu[\tilde g_{\mu\nu}],
\end{equation}
with $S_\nu[\tilde g_{\mu\nu}]$ representing the action for the neutrino, in which all contractions are made with a conformal metric 
$\tilde{g}_{\mu\nu}=e^{2\beta\frac{\phi}{\mpl}}g_{\mu\nu}$, where $\beta$ is a dimensionless constant. At the level of the  field theory, this is equivalent to expanding out the terms in the action to obtain 
\begin{equation}
\label{eq:act2}
S=\int d^4 x\sqrt{-g}\left[\frac{\mpl^2}{2} R(g)-\frac12\nabla_\mu\phi\nabla^\mu\phi-V(\phi)+i\bar\nu\gamma^\mu\nabla_\mu\nu+m_\nu\left(1+\beta\frac{\phi}{\mpl}+\cdots\right)\bar\nu\nu\right],
\end{equation}
where now all contractions are made just with $g_{\mu\nu}$. 

The scalar field equation of motion is then
\begin{equation}\label{eq:eom1}
\ddot{\phi}+3H\dot\phi+\frac{\partial V_{\rm eff}}{\partial \phi}=0,
\end{equation}
where we have specialized to a Friedmann-Robertson-Walker metric, with cosmic time, and where we have absorbed the coupling to neutrinos into an effective potential defined by
\begin{equation}
\label{eq:effpot}
V_{\rm eff}(\phi)\equiv V(\phi)-\beta\Theta(\nu)\frac{\phi}{\mpl}.
\end{equation}
Here $\Theta(\nu)=g_{\mu\nu}\Theta(\nu)^{\mu\nu}$ denotes the trace of the energy-momentum tensor for neutrinos, which is given by the standard expression 
\begin{equation}
\Theta(\nu)=-\rho_\nu+3P_\nu=-\frac{g_\nu T_\nu^4}{2\pi^2}\tau\left(\frac{m_\nu}{T_\nu}\right); \ \ \ \ \ \ \ \tau(x)=x^2\int_x^\infty\frac{\left(u^2-x^2\right)^{\frac12}}{e^u+1} \ du\label{eq:tauint},
\end{equation}
with $g_\nu$ the neutrino degeneracy and $T_\nu$ the neutrino temperature. 

Now, when $x\gg1$ we may neglect the neutrino mass and the neutrino equation of state is $P\approx\rho/3$, so that the integral $\tau(x)$ is approximately zero. Similarly, this integral is also approximately zero when $x\ll1$, due to Boltzmann suppression. Thus, the integral is only appreciably non-zero for $x\approx 1$ (where it is of order unity), corresponding to $T_\nu \approx m_\nu$. Therefore, as long as we avoid the parameter regime of regular early dark energy models, in which the mass of the EDE field is extremely small, the neutrinos contribute a forcing term to the scalar equation of motion that kicks the EDE scalar out of its minimum and up its potential when the temperature reaches $T_\nu\sim m_\nu$ (see, e.g.~\cite{Brax:2004qh}). 

To obtain an approximate idea of the magnitude of this kick, we can relate the neutrino temperature to the Hubble expansion via $3H^2\mpl^2=\pi^2/30 g_\star(T_\gamma)T_\gamma^4$, and assume that $T_\nu=m_\nu$ at a time $t_k$. Then, approximating the integral by $\tau(x)\approx 7\delta(t-t_k)/8H$, which injects energy over a Hubble time, we can neglect the contribution from the potential and perform the integral. It is easy to see, in this approximation~\cite{Sakstein:2019fmf}, that $\phi$ is displaced by
\begin{equation}\label{eq:phiin}
\phi_k\approx -0.03\beta\mpl.
\end{equation}
This result (which can be refined numerically) captures the main idea of $\nu$-assisted EDE, that the neutrino kick yields a natural initial condition for the EDE scalar, due to which it begins to roll a little before matter-radiation equality without us needing to fine-tune the mass to match the Hubble parameter at this epoch. The simple model presented above has a number of issues, particularly that it ignores the complexity of the behavior of neutrinos at higher temperatures and the behavior of the effective field theory of EDE. To address these issues it is useful to deal with a somewhat more complex setup.

\subsection{More realistic Models}
Going a step beyond the unrealistic and overly-simple model above, the next step that one can take is to work with the minimal phenomenologically viable, theoretically self-consistent model. This consists of a scalar coupled to a single massive neutrino species, without specifying whether the neutrinos are Dirac or Majorana, with more general couplings. This action is:
\begin{eqnarray}
\label{eq:act3}
S&=&\int d^4 x\sqrt{-g}\left[\frac{\mpl^2}{2} R(g)-\frac12\nabla_\mu\phi\nabla^\mu\phi-\frac12m^2\phi^2-\frac{\lambda}{4}\phi^4\right]+S_{\tilde{\nu}}[A^2(\phi) g_{\mu\nu}]\\\nonumber
&=&\int d^4 x\sqrt{-g}\left[\frac{\mpl^2}{2} R(g)-\frac12\nabla_\mu\phi\nabla^\mu\phi-\frac12m^2\phi^2-\frac{\lambda}{4}\phi^4+i\bar\nu\gamma^\mu\overset{\text{$\leftrightarrow$}}{\nabla}_\mu\nu-m_\nu A(\phi)\bar\nu\nu\right],
\end{eqnarray}
where we have again redefined the neutrino field as ${\nu}=A^{3/2}(\phi)\tilde{\nu}$. Two important things to note about this generalization are that we have included a mass term for the scalar field (since effective field theory dictates that we should write down the most general potential with a $\mathbb{Z}_2$ symmetry up to mass dimension four terms; more about this later), and we have generalized the coupling function, for similar reasons. 

The $\phi^4$ term provides an excellent fit to cosmological data when the mass is neglected \cite{Agrawal:2019lmo}. Adding a mass term leads to severe constraints from supernovae measurements of the late-time cosmic expansion rate but, from an EFT perspective, setting it to zero is unjustified. Indeed, if one imagines that this is a low energy EFT that arises from integrating out heavy degrees of freedom, then all operators compatible with the $\mathbb{Z}_2$ symmetry are expected to be present. Furthermore, once present, this mass is subject to quantum corrections from neutrino loops that arise from operators generated by the conformal coupling. It is presently unknown how large a mass is allowed by the data, and so a question of key importance in this work will be how small a choice we may make for the mass without spoiling the radiative stability of the model.

The generalization of the coupling function also opens up an opportunity for us to address a central shortcoming of the original simple model. Since $|\phi_{\rm min}|$ is an increasing function of redshift, at early times the $\phi^4$ term will dominate over the mass term, which can then be neglected. If we choose $A(\phi)\sim\exp(\beta\phi/\mpl)$ then the effective potential at early times is
\begin{equation}
    V_{\rm eff}(\phi)=V(\phi)-\frac{\beta \phi}{\mpl}(3{P}_\nu-{\rho}_\nu),
\end{equation}
which is minimized by
\begin{equation}\label{eq:phimagmin}
    |\phi_{\rm min}|=\left[\frac{\beta}{\lambda\mpl}\left(3{P}_\nu-{\rho}_\nu\right)\right]^{\frac{1}{3}}.
\end{equation}
Note that this quantity also grows with redshift, and hence that an exponential coupling will ultimately run afoul of EFT considerations,  requiring a UV-completion whose details will be important to the physics of the early universe, and calling into question some of our earlier conclusions. However, if we make the assumption that when the series is resummed, the coupling function has a minimum at some $\phi=-\bar\phi$ ($\bar\phi>0$) (similar to what happens in the strong coupling limit of string theory \cite{Damour:1994zq,Gasperini:2001pc,Brax:2010gi}), then the effective potential becomes
\begin{equation}
\label{eq, vefflogA}
    V_{\rm eff}(\phi)=V(\phi)+(\rho_\nu-3P_\nu)\ln[A(\phi)].
\end{equation}
Now, at high redshifts $(\rho_\nu-3P_\nu) (d\ln A/ d\phi) \gg V_{,\phi}(\phi)$ and we can ignore the potential, with the field minimizing the coupling function. As the neutrinos redshift, the potential becomes increasingly important and the minimum of the effective potential moves, but since $|\phi_{\rm min}|\le\bar\phi$, taking $\beta\bar\phi/\mpl<1$ is sufficient to ensure the evolution never leaves the range of validity of the EFT.  We can therefore write
\begin{equation}
\label{eq,Acoupling}
    A(\phi)=1-\frac{A_2}{2}\bar\phi^2+\frac{A_2}{2}\left(\phi+\bar\phi\right)^2+\cdots \ ,
\end{equation}
where the constant term is chosen to ensure that $A(\phi)=1+\beta\phi/\mpl+\cdots$. Expanding, this yields
 \begin{equation}
\label{eq,effA}
    A(\phi)= 1 +\frac{\beta\phi}{\mpl}+\frac{\beta}{2\mpl}\frac{\phi^2}{\bar\phi} \ ,
\end{equation}
with $\beta\equiv A_2 \mpl \bar{\phi}$. We can then safely neglect any additional terms in~\eqref{eq,Acoupling} since $\phi_{\rm min}\le\bar\phi$ for all of classical cosmological history, and the region of field space where these terms become important is thus never relevant.

\subsection{Quantum Corrections}

As we have pointed out, EFT considerations dictate that we must include a mass term for the scalar field. This is somewhat worrying empirically, since if the mass of the EDE field is too large since it will act as an additional dark matter component, which risks being incompatible with observations. However, we cannot choose the mass to be too small because of theoretical considerations - the mass should be radiatively stable to quantum corrections from the coupling between the scalar and the neutrino. It is therefore crucial to check whether there is a viable window with a sufficiently small, radiatively stable scalar mass.

We first expand around the background value of the EDE field $\phi = \phi_{\rm min} + \varphi$. At early times, $ \phi_{\rm min} = -\bar\phi$, while at late times this no longer holds. Expanding Equation~\eqref{eq:act3} and using Equation~\eqref{eq,effA}, we see that the bare mass for perturbations of the EDE field is
\begin{equation}
    m_{\rm bare}^2=m^2+3\lambda \phi_{\rm min}^2 \ , \label{baremass}
    \end{equation}
and the coupling to neutrinos is
\begin{equation}
 \left[ \frac{\beta}{\mpl} \left(1+ \frac{\phi_{\rm min}}{\bar \phi} \right) \varphi +\frac{\beta }{2\mpl \bar \phi} \varphi^2  \right] m_\nu \bar \nu \nu .
\end{equation}
Note also that there are now new cubic and quartic self-interactions of the field $\varphi$, which will induce 1-loop quantum corrections to the EDE mass. In~\cite{CarrilloGonzalez:2020oac} we computed all such corrections, demonstrating that all such corrections are negligible.

Of course, there are also corrections arising from the coupling to neutrinos. Again, in~\cite{CarrilloGonzalez:2020oac} one can see a detailed calculation showing that, in the specific case of Dirac neutrinos, in order to have a radiatively stable theory in which the loop corrections are smaller than the tree level result, it is sufficient to require that
\begin{equation} \label{mass}
    m^2+12\lambda\phi_{\rm min}^2 > \frac{\beta^2}{4\pi^2}\left(\frac{m_{\nu}}{\mpl}\right)^2m_\nu^2 \ .
\end{equation}

\subsection{Cosmological Evolution}

A qualitative description of the evolution of the EDE field is that at high redshifts, $(\rho_\nu-3P_\nu)\gg m^2\phi^2,\, \lambda\phi^4$, and the effective potential is minimized by $\phi=-\bar\phi$, which minimizes $\ln A(\phi)$. The field is therefore fixed motionless at this minimum. this yields robust initial conditions $\phi_{\rm in}=\bar\phi$, $\dot\phi_{\rm{in}}=0$. As $(3P_\nu-\rho_\nu)$ redshifts, the contribution from the quartic part of the scalar potential is no longer negligible, and the minimum begins to evolve adiabatically away from $\bar\phi$ towards zero. This happens when
\begin{equation}
    \lambda{\bar{\phi}}^3\sim \frac{\beta}{\mpl}\left(3P_\nu-\rho_\nu\right).
\end{equation}
After this, the higher-order terms in the expansion of the coupling function are irrelevant, and the evolution proceeds until $T_\nu\sim m_\nu$, when the neutrinos become non-relativistic and inject energy into the scalar, resulting in the kick feature in $\Omega_\phi$. This is the beginning of the EDE phase, responsible for resolving the Hubble tension. The kick is transient, and afterwards the field once again tracks the minimum of the effective potential, decreasing towards zero as the neutrino density redshifts. At some point the mass term dominates, and the minimum of the effective potential is
\begin{equation}
\label{eq,phiminmatter}
    \phi_{\rm min}=-\left(\frac{\beta\rho_\nu}{\mpl m^2}\right),
\end{equation}
where we have taken $P_\nu\ll\rho_\nu$ corresponding to $T_\nu\ll m_\nu$. 

During the matter dominated era, the field can be decomposed into an adiabatic component that tracks $\phi_{\rm min}$, and a rapidly-varying component $\delta \phi$ corresponding to oscillations about the (time-dependent) minimum. The adiabatic component acts as an additional dark energy component, but it is easy to show that for the required values of the parameters it constitutes a negligible amount of the observed dark energy, and a negligible correction to its equation of state.

More importantly, the rapidly-varying component executes oscillations around the nearly quadratic potential
\begin{equation}
    V(\delta\phi) = \frac{\lambda}{4} \delta \phi^4 +\lambda \delta \phi^3 \phi_{\rm min}+\frac{3\lambda}{2}\delta \phi^2\phi_{\rm min}^2+ \frac{m^2}{2}\delta \phi^2.  
\end{equation} 
Since the cross terms and the quartic term are negligible, the equation of motion becomes
\begin{equation}
\label{eq,deltaphieom}
    \ddot{\delta\phi}+3 H \dot{\delta\phi} +m^{2} \delta\phi=0 \ ,
\end{equation}
with solution  $\delta\phi = \frac{\delta\phi_0}{a^{3/2}} \cos(m t + \alpha)$, and corresponding fractional energy 
\begin{equation}
    \Omega_{\delta \phi} \equiv \frac{\frac{1}{2}\dot{\delta\phi}^2 + V(\delta\phi)}{3H^2\mpl^2}
\end{equation}
decaying as $a^{-3}$. This component therefore represents an additional contribution to the matter density. This is a source of tension with the data, and is the motivation behind neglecting a scalar mass term in earlier EDE models. An important question for all EDE models, including ours, is whether they can include a mass that is large enough that they are radiatively stable, but small enough that they are not in tension with the data. 

In~\cite{CarrilloGonzalez:2020oac} we numerically solved the full system of equations, including all other components of the energy budget of the universe. The results quantitatively verified the above expectations and allowed us to explore the parameter space.  We performed a thorough exploration of the correlation between the magnitude of the kick and the model parameters ( the mass of the scalar field $m$, the quartic self-coupling $\lambda$, and the coupling to neutrinos $\beta$). We also allowed the neutrino mass $m_\nu$ to vary modestly from the best-fit $\Lambda$CDM value obtained from Planck data, and identified those regions of parameter space where the kick magnitude is the largest around the epoch of matter-radiation equality. However, a full analysis to understand whether there exist viable regions of parameter space that resolve the Hubble tension is not yet completed, since it requires a significant modification of existing Boltzmann codes to allow for, in particular, that the Boltzmann hierarchy for massive neutrinos will be modified due to the non-minimal coupling to the scalar (see e.g.~\cite{Oldengott:2014qra}).   
 
\section{Chameleon EDE}

We now briefly  turn to a second way~\cite{Karwal:2021vpk} of addressing the fine-tuning problems of early dark energy by coupling to other fields. This time, we will be concerned with coupling between EDE and dark matter. Similarly to above, we consider an EDE field that conformally couples to dark matter, and refer to this model as {\it chameleon EDE}, since this coupling is inspired by chameleon models of dark energy~\cite{Khoury:2003aq, Khoury:2003rn}. The coupling to dark matter is important here since it allows us to trigger the dynamics of EDE when matter becomes the dominant component of the Universe at $z=z_{\rm eq}$. 

Consider the action
\begin{eqnarray} \label{Eq:ChameleonAction}
    S &= & \int {\rm d}^4 x\sqrt{-g} \left[ \frac{\mpl}{2}\mathcal{R} -\frac{1}{2}(\nabla \phi)^2 -V(\phi) \right] \nonumber \\ 
    & &+S_{\dm}[\phi_\dm, \tilde{g}_{\mu\nu} ] +S_{\rm m}[\phi_{\rm m}, g_{\mu\nu} ] \,,
\end{eqnarray}
where $m$ denotes all cosmic matter except dark matter, and the two metrics, $g_{\mu \nu}$ and $\tilde{g}_{\mu \nu}$, are again related through some arbitrary function $A(\phi)$ of the scalar $\phi$ by 
\begin{equation} \label{Eq:ConformalMetric}
    \tilde{g}_{\mu\nu} = A^2(\phi) g_{\mu\nu} \,.
\end{equation}
Here, photons and baryons move on geodesics in the Einstein frame with metric $g_{\mu \nu}$, but DM motion is also influenced by acceleration due to the scalar field. Alternatively, one can say that DM moves on geodesics in the Jordan frame, with metric $\tilde{g}_{\mu \nu}$, but photons and baryons are accelerated in that frame.
We write quantities defined in the Jordan frame with a tilde and quantities defined in the Einstein frame without a tilde, and generally work and solve equations in the Einstein frame for convenience.

We assume that different matter species have stress-energy tensors of perfect fluids, in their respective geodesic frames.
So, for the DM component we have 
\begin{equation}
    \tilde{T}_{\mu\nu} = (\tilde{\rho} +\tilde{P})\tilde{u}_\mu\tilde{u}_\nu +\tilde{P}\tilde{g}_{\mu\nu} \,.
\end{equation}

At the background level in the Einstein frame, the Friedmann equation is
\begin{equation}
    3 \mathcal{H}^2 \mpl^2
    = \frac{1}{2}{\phi}'^2 + a^2 V(\phi)
    + \tilde{\rho}_{\dm}A^4(\phi) a^2 
    + \rho_m a^2 + \rho_\Lambda a^2\ ,
\end{equation}
where $\mathcal{H} = {a'}/a$ is the Hubble parameter in conformal time, and primes represent derivatives with respect to conformal time. The equation of motion of the scalar is 
\begin{equation}
    {\phi}'' + 2\mathcal{H}{\phi}' 
    = - a^2 V_{,\phi}
    - a^2 A_{,\phi}A^3(\phi) \tilde{\rho}_{\dm} \,.
\end{equation}
The continuity equation for dark matter yields
\begin{equation}
    \tilde{\rho}_\dm 
    = \tilde{\rho}_\dm^{0} a^{-3}\left(\frac{A_0}{ A} \right)^3 \,,
\end{equation}
where a subscript or superscript $0$ represents the value of a quantity today, at $a=1$. It is also convenient to define an auxiliary quantity, $\rho_{\dm}$, as $\rho_\dm \equiv \tilde{\rho}_\dm A^3$, which now dilutes like standard CDM as $\rho_{\dm} \propto a^{-3}$. 
The effective potential $V_{\rm eff}(\phi)$ of the scalar in the Einstein frame can then be written as 
\begin{equation}
    V_{\rm eff}(\phi)
    = V(\phi)
    + A(\phi) \rho_{\dm} \,.
\end{equation}

The Hubble constraint equation evaluated today reads
\begin{equation}
    1 = \Omega_m^0 +A(\phi_0) \Omega_{\dm,0} +\Omega_\phi^0 +\Omega_\Lambda^0 \, ,
\end{equation}
where  the physical meaning of $\Omega_{\dm,0}$ is not the usual one and instead should just be considered an auxiliary variable. 
The relative DM gravitational density is given by $\Omega_{\dm,0} A(\phi_0)$, while the rest frame density is given by $\Omega_{\dm,0} A^{-3}(\phi_0)$. Hence, at any point in time, the contribution of DM to the total energy budget of the Universe is $\rho_{\dm}A(\phi)$, which can be interpreted as a modulation of the mass of the DM particle. 

\subsection{Initial conditions}
It is worth briefly discussing the initial conditions for the cosmological evolution of this system. Initially, no matter what the potential is, the coupled scalar field is dominated by the kinetic energy ${\phi}'^2/2a^2$ 
and $\phi$ rolls down the time-varying potential $A(\phi)\rho_\dm$. 
This holds for several order of magnitude in $A(\phi_{\rm i})$, as $\rho_\dm$ is very large at early times, and most reasonable choices of $\phi_{\rm i} \sim \mpl$ and therefore $V(\phi_{\rm i}) \ll \rho_\dm(a_{\rm i})$ are not too large. 

During this time, there is little change in $\phi$ and its potential energy. 
At initial times, assuming ${\phi}''_{\rm i}\rightarrow 0$, we set
\begin{eqnarray}
    2\mathcal{H}{\phi'_{\rm i}} 
    & \simeq & - a^2 V_{,\phi}
    - a^2 A_{,\phi} \rho_{\dm} \nonumber\\
    & \Rightarrow & {\phi}'_{\rm i} \simeq -\frac{a^2}{2\mathcal{H}} \big( V_{, \phi} + A_{,\phi}\rho_{\dm} \big) \,,
    \label{eq:phi_prime_init_cond}
\end{eqnarray}
which recovers the uncoupled regime as $A \rightarrow 0$. 
As ${\phi}'_{\rm i}$ is set deep in radiation domination, if the right-hand side is dictated by the DM term, ${\phi}'_{\rm i}$ is roughly constant, independent of $a$, validating setting ${\phi}''_{\rm i} = 0$. 
Then, $\rho_{\rm scf} \propto a^{-2}$ at early times, dominated by the kinetic energy of $\phi$. 

The initial field location $\phi_{\rm i}$, on the other hand, is not set by an attractor solution, but is an input parameter that controls the maximal fractional energy density $f_{\ede}$ in CEDE. 
There is also a degeneracy between $\phi_{\rm i}$ and $\Omega_{\dm,{\rm i}}$ wherein changing $\phi_{\rm i}$ simply rescales $\Omega_{\dm,{\rm i}}$. 

\subsection{Models of CEDE} \label{Sec:CEDEmodels}

With this general CEDE setup, we make the following choices of scalar field potential and form of the coupling 
\begin{eqnarray}
    A(\phi) &=& e^{\beta \phi/\mpl} \,\, {\rm and } \nonumber\\
    V(\phi) &=& \lambda \phi^4 \,,
\end{eqnarray}
where $\lambda$ and $\beta$ are dimensionless constants. 
Note that we have chosen a a $\phi^4$ potential to ensure that after it is active, EDE dilutes away at least as fast as radiation. Of course, since the model contains no symmetry forbidding a mass term, the rules of effective field theory dictate that we should, 
in principle, include one, since if we do not then such a term will be generated by radiative corrections. 
For $f_\ede \sim O(10\%)$ and $z_c \sim z_{\rm eq}$, the scalar field must be ultra-light, with~$m_{\phi}(\phi_{\rm i}) \sim \sqrt{\lambda} \phi_{\rm i} \sim 10^{-28}~{\rm eV}$. 
The scalar mass will receive radiative corrections, for example from loops of dark matter, of order 
\begin{equation}
\delta m_{\phi} \sim \beta \frac{\Lambda^2}{M_{\rm Pl}}\,.
\end{equation}
The cutoff~$\Lambda$ should at the very least be much greater than the dark matter particle mass~$m_{\rm dm}$, and hence radiative stability requires~$m_{\rm dm} \ll {\rm eV}$. This will be the case if dark matter is axion-like. (Similar considerations show that the quartic coupling is also radiatively stable for sufficiently light dark matter.) Thus the inclusion of a tree-level mass term, such that $V(\phi)$ becomes
\begin{equation}
    V(\phi) = \frac{1}{2}m^2 \phi^2 + \lambda \phi^4 \, ,
\end{equation}
with $m \lesssim 10^{-28}~{\rm eV}$ will have little impact on cosmology.

An uncoupled scalar field with a $\phi^4$ potential is initially frozen due to Hubble friction. 
When it begins to roll, its energy density dilutes $\propto a^{-4}$ when time-averaged over oscillations. 
Such a dissipation satisfies the EDE requirement of vanishing at late times and has already been explored in \cite{Agrawal:2019lmo}. 

At the background level, the introduction of the coupling to DM most noticeably modifies the early-time behavior of the scalar field, when the field is dominated by kinetic energy ${\phi}'^2/2a^2$, as described by Eq.~\eqref{eq:phi_prime_init_cond} with $\rho_{\rm scf}$ scaling as $a^{-2}$. 
Furthermore, as $\beta$ increases, the effective potential $V_{\rm eff}$ felt by the scalar differs from $V(\phi)$ to a greater extent, becoming increasingly asymmetric
\begin{equation}
    V_{\rm eff}(\phi) = \lambda \phi^4 
                        + e^{\beta \phi/\mpl} \rho_{\dm} \,.
    \label{eq:eff_pot}
\end{equation}
It is this direct coupling to DM energy density that offers the possibility of EDE dynamics being triggered by DM becoming the dominant component of the Universe. 

The subsequent evolution of the field can be divided into two distinct scenarios - one in which $\beta \phi_{\rm i}/\mpl < 0$ and another where $\beta \phi_{\rm i}/\mpl > 0$.
Let us assume $\phi_{\rm i} / \mpl >0 $. 
The more commonly explored chameleon dark energy case sets $\beta$ and $\phi_{\rm i}/\mpl$ with opposite signs. 
In this scenario, the two terms contributing to ${\phi}'_{\rm i}$ in Eq.~\eqref{eq:phi_prime_init_cond} have opposite signs, slowing down $|{\phi}'_{\rm i}|$ relative to the $\beta >0$ case. 
Note that $|{\phi}'_{\rm i}|$ is still larger than in the uncoupled case, as the DM term dominates the contribution to ${\phi}'_{\rm i}$.  
Moreover, ${\phi}'_{\rm i} > 0$ and $\phi$ is initially being kicked up its native potential to larger values, moving towards the minimum of its effective potential in Eq.~\eqref{eq:eff_pot}, higher than in the uncoupled case. 
Then, as $|\beta|$ increases, the field is at a higher point in its potential at early times, with greater energy density than in uncoupled EDE or $\beta > 0$ CEDE.

On the other hand, for $\beta \phi_{\rm i}/\mpl >0$, $|{\phi}'_{\rm i}|$ is larger than in the $\beta < 0$ CEDE case, and is negative. 
Hence, $\phi$ is lower in its potential relative to the uncoupled or $\beta < 0$ CEDE cases at early times. 
Accordingly, as $\beta$ increases, CEDE has smaller $f_\ede$ in this case than in uncoupled EDE. 

As DM dilutes away, the native potential $V(\phi)$ of the scalar begins to dictate its dynamics. 
Both $\beta >0$ and $\beta<0$ scenarios may then become Hubble frozen for some decades in redshift.
The value of $\beta$ controls the duration of this Hubble-frozen period, with higher $\beta$ leading to a smaller or no frozen window.
For smaller values of $\beta$, it is $\lambda_{\rm scf}$ that controls the redshift $z_c$ at which $f_\ede$ peaks, similar to uncoupled EDE. 

The scalar then begins to roll and oscillate in $V_{\rm eff}$ about a new, time-varying minimum shifted from 0, defined by the solution to 
\begin{equation}
    4\lambda \phi_{\rm min}^3 = - \frac{\beta}{\mpl}e^{\beta \phi_{\rm min}/\mpl} \rho_{\dm} \,.
\end{equation}
Although the time-averaged density of the field falls as $a^{-4}$ during this period, $\phi$ undergoes asymmetric oscillations about this new minimum. 
This shifts the odd (even) peaks in $f_\ede$ to lower (higher) energy density than in the symmetric-potential uncoupled EDE case for $\beta > 0$ and vice versa for $\beta < 0$. 

In~\cite{Karwal:2021vpk} we performed a first analysis of the most promising parts of parameter space for CEDE models to succeed in solving the Hubble tension. Rather than reproducing all that in this talk, I have just highlighted the interesting features of such models and the expectation of improvement over an uncoupled EDE model. However, as in the case of $\nu$EDE, to fully and robustly understand the impact of data on CEDE models, it is necessary to perform Markov chain Monte Carlo (MCMC) searches in parameter space. This work is now underway.

\section{Summary}
In this presentation, I have very briefly discussed two different theoretical approaches to addressing fine-tuning issues in EDE models. In both cases we exploit a coupling of the EDE field to physics that already has a feature that takes place approximately at the time of matter-radiation equality. The main idea is for this physics to act as a trigger for the onset of early dark energy. 

In the first of these approaches~\cite{Sakstein:2019fmf,CarrilloGonzalez:2020oac}, termed {\it Neutrino-Assisted Early Dark Energy}, we couple the EDE field to neutrinos, activating the epoch of early dark energy as neutrinos transition from being relativistic to non-relativistic. This is to be compared to the second approach~\cite{Karwal:2021vpk} in which we couple the EDE field to dark matter, so that early dark energy becomes an important feature when dark matter becomes the dominant component of the Universe. This class of models is termed {\it Chameleon Early Dark Energy}.

I have described the background evolution in each case, and have explained how the models avoid many of the fine-tuning issues, including those arising from effective field theory considerations, present in uncoupled EDE models. Preliminary analysis indicates that these models may provide interesting alternative solutions to the Hubble tension, but themselves face strict tests from current and upcoming cosmological datasets. In order to understand whether either of these models is viable, a full MCMC analysis of both is needed. While this work is underway, I did not present any such results in this talk, but hope that these will be available in upcoming publications.

\section*{Acknowledgements}
I would like to thank the organizers of the 15th Frontiers of Fundamental Physics conference for the invitation to deliver this talk, and for curating such an interesting roster of other presentations. I am extremely grateful to Tanvi Karwal for providing comments on this manuscript. I would also like to thank Mariana Carrillo Gonzalez, Bhuvnesh Jain, Tanvi Karwal, Justin Khoury, Qiuyue Liang, Marco Raveri, and Jeremy Sakstein for thoroughly enjoyable collaborations. This work was supported in part by US Department of Energy (HEP) Award DE- SC0013528, and by NASA ATP grant 80NSSC18K0694.

\bibliography{ffp}

\begin{thebibliography}{37}
\expandafter\ifx\csname natexlab\endcsname\relax\def\natexlab#1{#1}\fi
\expandafter\ifx\csname bibnamefont\endcsname\relax
  \def\bibnamefont#1{#1}\fi
\expandafter\ifx\csname bibfnamefont\endcsname\relax
  \def\bibfnamefont#1{#1}\fi
\expandafter\ifx\csname citenamefont\endcsname\relax
  \def\citenamefont#1{#1}\fi
\expandafter\ifx\csname url\endcsname\relax
  \def\url#1{\texttt{#1}}\fi
\expandafter\ifx\csname urlprefix\endcsname\relax\def\urlprefix{URL }\fi
\providecommand{\bibinfo}[2]{#2}
\providecommand{\eprint}[2][]{\url{#2}}

\bibitem[{\citenamefont{Aghanim et~al.}(2020)}]{Aghanim:2018eyx}
\bibinfo{author}{\bibfnamefont{N.}~\bibnamefont{Aghanim}} \bibnamefont{et~al.}
  (\bibinfo{collaboration}{Planck}), \bibinfo{journal}{Astron. Astrophys.}
  \textbf{\bibinfo{volume}{641}}, \bibinfo{pages}{A6} (\bibinfo{year}{2020}),
  \bibinfo{note}{[Erratum: Astron.Astrophys. 652, C4 (2021)]},
  \eprint{1807.06209}.

\bibitem[{\citenamefont{Riess et~al.}(2021)}]{Riess:2021jrx}
\bibinfo{author}{\bibfnamefont{A.~G.} \bibnamefont{Riess}} \bibnamefont{et~al.}
  (\bibinfo{year}{2021}), \eprint{2112.04510}.

\bibitem[{\citenamefont{Freedman et~al.}(2020)\citenamefont{Freedman, Madore,
  Hoyt, Jang, Beaton, Lee, Monson, Neeley, and Rich}}]{Freedman:2020dne}
\bibinfo{author}{\bibfnamefont{W.~L.} \bibnamefont{Freedman}},
  \bibinfo{author}{\bibfnamefont{B.~F.} \bibnamefont{Madore}},
  \bibinfo{author}{\bibfnamefont{T.}~\bibnamefont{Hoyt}},
  \bibinfo{author}{\bibfnamefont{I.~S.} \bibnamefont{Jang}},
  \bibinfo{author}{\bibfnamefont{R.}~\bibnamefont{Beaton}},
  \bibinfo{author}{\bibfnamefont{M.~G.} \bibnamefont{Lee}},
  \bibinfo{author}{\bibfnamefont{A.}~\bibnamefont{Monson}},
  \bibinfo{author}{\bibfnamefont{J.}~\bibnamefont{Neeley}}, \bibnamefont{and}
  \bibinfo{author}{\bibfnamefont{J.}~\bibnamefont{Rich}}
  (\bibinfo{year}{2020}), \eprint{2002.01550}.

\bibitem[{\citenamefont{Wong et~al.}(2020)}]{Wong:2019kwg}
\bibinfo{author}{\bibfnamefont{K.~C.} \bibnamefont{Wong}} \bibnamefont{et~al.},
  \bibinfo{journal}{Mon. Not. Roy. Astron. Soc.}
  \textbf{\bibinfo{volume}{498}}, \bibinfo{pages}{1420} (\bibinfo{year}{2020}),
  \eprint{1907.04869}.

\bibitem[{\citenamefont{Birrer et~al.}(2020)}]{Birrer:2020tax}
\bibinfo{author}{\bibfnamefont{S.}~\bibnamefont{Birrer}} \bibnamefont{et~al.},
  \bibinfo{journal}{Astron. Astrophys.} \textbf{\bibinfo{volume}{643}},
  \bibinfo{pages}{A165} (\bibinfo{year}{2020}), \eprint{2007.02941}.

\bibitem[{\citenamefont{Verde et~al.}(2019)\citenamefont{Verde, Treu, and
  Riess}}]{Verde:2019ivm}
\bibinfo{author}{\bibfnamefont{L.}~\bibnamefont{Verde}},
  \bibinfo{author}{\bibfnamefont{T.}~\bibnamefont{Treu}}, \bibnamefont{and}
  \bibinfo{author}{\bibfnamefont{A.~G.} \bibnamefont{Riess}},
  \bibinfo{journal}{Nature Astron.} \textbf{\bibinfo{volume}{3}},
  \bibinfo{pages}{891} (\bibinfo{year}{2019}), \eprint{1907.10625}.

\bibitem[{\citenamefont{Fitzpatrick et~al.}(2000)\citenamefont{Fitzpatrick,
  Ribas, Guinan, DeWarf, Maloney, and Massa}}]{Fitzpatrick:2000hh}
\bibinfo{author}{\bibfnamefont{E.~L.} \bibnamefont{Fitzpatrick}},
  \bibinfo{author}{\bibfnamefont{I.}~\bibnamefont{Ribas}},
  \bibinfo{author}{\bibfnamefont{E.~F.} \bibnamefont{Guinan}},
  \bibinfo{author}{\bibfnamefont{L.~E.} \bibnamefont{DeWarf}},
  \bibinfo{author}{\bibfnamefont{F.~P.} \bibnamefont{Maloney}},
  \bibnamefont{and} \bibinfo{author}{\bibfnamefont{D.}~\bibnamefont{Massa}}
  (\bibinfo{year}{2000}), \eprint{astro-ph/0010526}.

\bibitem[{\citenamefont{Benedict et~al.}(2007)\citenamefont{Benedict, McArthur,
  Feast, Barnes, Harrison, Patterson, Menzies, Bean, and
  Freedman}}]{Benedict:2006cp}
\bibinfo{author}{\bibfnamefont{G.~F.} \bibnamefont{Benedict}},
  \bibinfo{author}{\bibfnamefont{B.~E.} \bibnamefont{McArthur}},
  \bibinfo{author}{\bibfnamefont{M.~W.} \bibnamefont{Feast}},
  \bibinfo{author}{\bibfnamefont{T.~G.} \bibnamefont{Barnes}},
  \bibinfo{author}{\bibfnamefont{T.~E.} \bibnamefont{Harrison}},
  \bibinfo{author}{\bibfnamefont{R.~J.} \bibnamefont{Patterson}},
  \bibinfo{author}{\bibfnamefont{J.~W.} \bibnamefont{Menzies}},
  \bibinfo{author}{\bibfnamefont{J.~L.} \bibnamefont{Bean}}, \bibnamefont{and}
  \bibinfo{author}{\bibfnamefont{W.~L.} \bibnamefont{Freedman}},
  \bibinfo{journal}{Astron. J.} \textbf{\bibinfo{volume}{133}},
  \bibinfo{pages}{1810} (\bibinfo{year}{2007}), \bibinfo{note}{[Erratum:
  Astron.J. 133, 2980 (2007)]}, \eprint{astro-ph/0612465}.

\bibitem[{\citenamefont{Humphreys et~al.}(2013)\citenamefont{Humphreys, Reid,
  Moran, Greenhill, and Argon}}]{Humphreys:2013eja}
\bibinfo{author}{\bibfnamefont{E.~M.~L.} \bibnamefont{Humphreys}},
  \bibinfo{author}{\bibfnamefont{M.~J.} \bibnamefont{Reid}},
  \bibinfo{author}{\bibfnamefont{J.~M.} \bibnamefont{Moran}},
  \bibinfo{author}{\bibfnamefont{L.~J.} \bibnamefont{Greenhill}},
  \bibnamefont{and} \bibinfo{author}{\bibfnamefont{A.~L.} \bibnamefont{Argon}},
  \bibinfo{journal}{Astrophys. J.} \textbf{\bibinfo{volume}{775}},
  \bibinfo{pages}{13} (\bibinfo{year}{2013}), \eprint{1307.6031}.

\bibitem[{\citenamefont{Efstathiou}(2014)}]{Efstathiou:2013via}
\bibinfo{author}{\bibfnamefont{G.}~\bibnamefont{Efstathiou}},
  \bibinfo{journal}{Mon. Not. Roy. Astron. Soc.}
  \textbf{\bibinfo{volume}{440}}, \bibinfo{pages}{1138} (\bibinfo{year}{2014}),
  \eprint{1311.3461}.

\bibitem[{\citenamefont{Rigault et~al.}(2015)}]{Rigault:2014kaa}
\bibinfo{author}{\bibfnamefont{M.}~\bibnamefont{Rigault}} \bibnamefont{et~al.},
  \bibinfo{journal}{Astrophys. J.} \textbf{\bibinfo{volume}{802}},
  \bibinfo{pages}{20} (\bibinfo{year}{2015}), \eprint{1412.6501}.

\bibitem[{\citenamefont{Becker et~al.}(2015)\citenamefont{Becker, Desmond,
  Rozo, Marshall, and Rykoff}}]{Becker:2015nya}
\bibinfo{author}{\bibfnamefont{M.~R.} \bibnamefont{Becker}},
  \bibinfo{author}{\bibfnamefont{H.}~\bibnamefont{Desmond}},
  \bibinfo{author}{\bibfnamefont{E.}~\bibnamefont{Rozo}},
  \bibinfo{author}{\bibfnamefont{P.}~\bibnamefont{Marshall}}, \bibnamefont{and}
  \bibinfo{author}{\bibfnamefont{E.~S.} \bibnamefont{Rykoff}}
  (\bibinfo{year}{2015}), \eprint{1507.07523}.

\bibitem[{\citenamefont{Shanks et~al.}(2019)\citenamefont{Shanks, Hogarth, and
  Metcalfe}}]{Shanks:2018rka}
\bibinfo{author}{\bibfnamefont{T.}~\bibnamefont{Shanks}},
  \bibinfo{author}{\bibfnamefont{L.}~\bibnamefont{Hogarth}}, \bibnamefont{and}
  \bibinfo{author}{\bibfnamefont{N.}~\bibnamefont{Metcalfe}},
  \bibinfo{journal}{Mon. Not. Roy. Astron. Soc.}
  \textbf{\bibinfo{volume}{484}}, \bibinfo{pages}{L64} (\bibinfo{year}{2019}),
  \eprint{1810.02595}.

\bibitem[{\citenamefont{Riess et~al.}(2018)\citenamefont{Riess, Casertano,
  Kenworthy, Scolnic, and Macri}}]{Riess:2018kzi}
\bibinfo{author}{\bibfnamefont{A.~G.} \bibnamefont{Riess}},
  \bibinfo{author}{\bibfnamefont{S.}~\bibnamefont{Casertano}},
  \bibinfo{author}{\bibfnamefont{D.}~\bibnamefont{Kenworthy}},
  \bibinfo{author}{\bibfnamefont{D.}~\bibnamefont{Scolnic}}, \bibnamefont{and}
  \bibinfo{author}{\bibfnamefont{L.}~\bibnamefont{Macri}}
  (\bibinfo{year}{2018}), \eprint{1810.03526}.

\bibitem[{\citenamefont{Kenworthy et~al.}(2019)\citenamefont{Kenworthy,
  Scolnic, and Riess}}]{Kenworthy:2019qwq}
\bibinfo{author}{\bibfnamefont{W.~D.} \bibnamefont{Kenworthy}},
  \bibinfo{author}{\bibfnamefont{D.}~\bibnamefont{Scolnic}}, \bibnamefont{and}
  \bibinfo{author}{\bibfnamefont{A.}~\bibnamefont{Riess}},
  \bibinfo{journal}{Astrophys. J.} \textbf{\bibinfo{volume}{875}},
  \bibinfo{pages}{145} (\bibinfo{year}{2019}), \eprint{1901.08681}.

\bibitem[{\citenamefont{Spergel et~al.}(2015)\citenamefont{Spergel, Flauger,
  and Hlo\v{z}ek}}]{Spergel:2013rxa}
\bibinfo{author}{\bibfnamefont{D.~N.} \bibnamefont{Spergel}},
  \bibinfo{author}{\bibfnamefont{R.}~\bibnamefont{Flauger}}, \bibnamefont{and}
  \bibinfo{author}{\bibfnamefont{R.}~\bibnamefont{Hlo\v{z}ek}},
  \bibinfo{journal}{Phys. Rev. D} \textbf{\bibinfo{volume}{91}},
  \bibinfo{pages}{023518} (\bibinfo{year}{2015}), \eprint{1312.3313}.

\bibitem[{\citenamefont{Karwal and Kamionkowski}(2016)}]{Karwal:2016vyq}
\bibinfo{author}{\bibfnamefont{T.}~\bibnamefont{Karwal}} \bibnamefont{and}
  \bibinfo{author}{\bibfnamefont{M.}~\bibnamefont{Kamionkowski}},
  \bibinfo{journal}{Phys. Rev. D} \textbf{\bibinfo{volume}{94}},
  \bibinfo{pages}{103523} (\bibinfo{year}{2016}), \eprint{1608.01309}.

\bibitem[{\citenamefont{Poulin et~al.}(2019)\citenamefont{Poulin, Smith,
  Karwal, and Kamionkowski}}]{Poulin:2018cxd}
\bibinfo{author}{\bibfnamefont{V.}~\bibnamefont{Poulin}},
  \bibinfo{author}{\bibfnamefont{T.~L.} \bibnamefont{Smith}},
  \bibinfo{author}{\bibfnamefont{T.}~\bibnamefont{Karwal}}, \bibnamefont{and}
  \bibinfo{author}{\bibfnamefont{M.}~\bibnamefont{Kamionkowski}},
  \bibinfo{journal}{Phys. Rev. Lett.} \textbf{\bibinfo{volume}{122}},
  \bibinfo{pages}{221301} (\bibinfo{year}{2019}), \eprint{1811.04083}.

\bibitem[{\citenamefont{Bernal et~al.}(2016)\citenamefont{Bernal, Verde, and
  Riess}}]{Bernal:2016gxb}
\bibinfo{author}{\bibfnamefont{J.~L.} \bibnamefont{Bernal}},
  \bibinfo{author}{\bibfnamefont{L.}~\bibnamefont{Verde}}, \bibnamefont{and}
  \bibinfo{author}{\bibfnamefont{A.~G.} \bibnamefont{Riess}},
  \bibinfo{journal}{JCAP} \textbf{\bibinfo{volume}{10}}, \bibinfo{pages}{019}
  (\bibinfo{year}{2016}), \eprint{1607.05617}.

\bibitem[{\citenamefont{Evslin et~al.}(2018)\citenamefont{Evslin, Sen, and
  Ruchika}}]{Evslin:2017qdn}
\bibinfo{author}{\bibfnamefont{J.}~\bibnamefont{Evslin}},
  \bibinfo{author}{\bibfnamefont{A.~A.} \bibnamefont{Sen}}, \bibnamefont{and}
  \bibinfo{author}{\bibnamefont{Ruchika}}, \bibinfo{journal}{Phys. Rev. D}
  \textbf{\bibinfo{volume}{97}}, \bibinfo{pages}{103511}
  (\bibinfo{year}{2018}), \eprint{1711.01051}.

\bibitem[{\citenamefont{Aylor et~al.}(2019)\citenamefont{Aylor, Joy, Knox,
  Millea, Raghunathan, and Wu}}]{Aylor:2018drw}
\bibinfo{author}{\bibfnamefont{K.}~\bibnamefont{Aylor}},
  \bibinfo{author}{\bibfnamefont{M.}~\bibnamefont{Joy}},
  \bibinfo{author}{\bibfnamefont{L.}~\bibnamefont{Knox}},
  \bibinfo{author}{\bibfnamefont{M.}~\bibnamefont{Millea}},
  \bibinfo{author}{\bibfnamefont{S.}~\bibnamefont{Raghunathan}},
  \bibnamefont{and} \bibinfo{author}{\bibfnamefont{W.~L.~K.} \bibnamefont{Wu}},
  \bibinfo{journal}{Astrophys. J.} \textbf{\bibinfo{volume}{874}},
  \bibinfo{pages}{4} (\bibinfo{year}{2019}), \eprint{1811.00537}.

\bibitem[{\citenamefont{Knox and Millea}(2020)}]{Knox:2019rjx}
\bibinfo{author}{\bibfnamefont{L.}~\bibnamefont{Knox}} \bibnamefont{and}
  \bibinfo{author}{\bibfnamefont{M.}~\bibnamefont{Millea}},
  \bibinfo{journal}{Phys. Rev. D} \textbf{\bibinfo{volume}{101}},
  \bibinfo{pages}{043533} (\bibinfo{year}{2020}), \eprint{1908.03663}.

\bibitem[{\citenamefont{Poulin et~al.}(2018)\citenamefont{Poulin, Smith, Grin,
  Karwal, and Kamionkowski}}]{Poulin:2018dzj}
\bibinfo{author}{\bibfnamefont{V.}~\bibnamefont{Poulin}},
  \bibinfo{author}{\bibfnamefont{T.~L.} \bibnamefont{Smith}},
  \bibinfo{author}{\bibfnamefont{D.}~\bibnamefont{Grin}},
  \bibinfo{author}{\bibfnamefont{T.}~\bibnamefont{Karwal}}, \bibnamefont{and}
  \bibinfo{author}{\bibfnamefont{M.}~\bibnamefont{Kamionkowski}},
  \bibinfo{journal}{Phys. Rev. D} \textbf{\bibinfo{volume}{98}},
  \bibinfo{pages}{083525} (\bibinfo{year}{2018}), \eprint{1806.10608}.

\bibitem[{\citenamefont{Smith et~al.}(2020)\citenamefont{Smith, Poulin, and
  Amin}}]{Smith:2019ihp}
\bibinfo{author}{\bibfnamefont{T.~L.} \bibnamefont{Smith}},
  \bibinfo{author}{\bibfnamefont{V.}~\bibnamefont{Poulin}}, \bibnamefont{and}
  \bibinfo{author}{\bibfnamefont{M.~A.} \bibnamefont{Amin}},
  \bibinfo{journal}{Phys. Rev. D} \textbf{\bibinfo{volume}{101}},
  \bibinfo{pages}{063523} (\bibinfo{year}{2020}), \eprint{1908.06995}.

\bibitem[{\citenamefont{Sakstein and Trodden}(2020)}]{Sakstein:2019fmf}
\bibinfo{author}{\bibfnamefont{J.}~\bibnamefont{Sakstein}} \bibnamefont{and}
  \bibinfo{author}{\bibfnamefont{M.}~\bibnamefont{Trodden}},
  \bibinfo{journal}{Phys. Rev. Lett.} \textbf{\bibinfo{volume}{124}},
  \bibinfo{pages}{161301} (\bibinfo{year}{2020}), \eprint{1911.11760}.

\bibitem[{\citenamefont{Carrillo~Gonz\'alez
  et~al.}(2021)\citenamefont{Carrillo~Gonz\'alez, Liang, Sakstein, and
  Trodden}}]{CarrilloGonzalez:2020oac}
\bibinfo{author}{\bibfnamefont{M.}~\bibnamefont{Carrillo~Gonz\'alez}},
  \bibinfo{author}{\bibfnamefont{Q.}~\bibnamefont{Liang}},
  \bibinfo{author}{\bibfnamefont{J.}~\bibnamefont{Sakstein}}, \bibnamefont{and}
  \bibinfo{author}{\bibfnamefont{M.}~\bibnamefont{Trodden}},
  \bibinfo{journal}{JCAP} \textbf{\bibinfo{volume}{04}}, \bibinfo{pages}{063}
  (\bibinfo{year}{2021}), \eprint{2011.09895}.

\bibitem[{\citenamefont{Aker et~al.}(2019)}]{Aker:2019uuj}
\bibinfo{author}{\bibfnamefont{M.}~\bibnamefont{Aker}} \bibnamefont{et~al.}
  (\bibinfo{collaboration}{KATRIN}), \bibinfo{journal}{Phys. Rev. Lett.}
  \textbf{\bibinfo{volume}{123}}, \bibinfo{pages}{221802}
  (\bibinfo{year}{2019}), \eprint{1909.06048}.

\bibitem[{\citenamefont{Karwal et~al.}(2022)\citenamefont{Karwal, Raveri, Jain,
  Khoury, and Trodden}}]{Karwal:2021vpk}
\bibinfo{author}{\bibfnamefont{T.}~\bibnamefont{Karwal}},
  \bibinfo{author}{\bibfnamefont{M.}~\bibnamefont{Raveri}},
  \bibinfo{author}{\bibfnamefont{B.}~\bibnamefont{Jain}},
  \bibinfo{author}{\bibfnamefont{J.}~\bibnamefont{Khoury}}, \bibnamefont{and}
  \bibinfo{author}{\bibfnamefont{M.}~\bibnamefont{Trodden}},
  \bibinfo{journal}{Phys. Rev. D} \textbf{\bibinfo{volume}{105}},
  \bibinfo{pages}{063535} (\bibinfo{year}{2022}), \eprint{2106.13290}.

\bibitem[{\citenamefont{Kamionkowski et~al.}(2014)\citenamefont{Kamionkowski,
  Pradler, and Walker}}]{Kamionkowski:2014zda}
\bibinfo{author}{\bibfnamefont{M.}~\bibnamefont{Kamionkowski}},
  \bibinfo{author}{\bibfnamefont{J.}~\bibnamefont{Pradler}}, \bibnamefont{and}
  \bibinfo{author}{\bibfnamefont{D.~G.~E.} \bibnamefont{Walker}},
  \bibinfo{journal}{Phys. Rev. Lett.} \textbf{\bibinfo{volume}{113}},
  \bibinfo{pages}{251302} (\bibinfo{year}{2014}), \eprint{1409.0549}.

\bibitem[{\citenamefont{Brax et~al.}(2004)\citenamefont{Brax, van~de Bruck,
  Davis, Khoury, and Weltman}}]{Brax:2004qh}
\bibinfo{author}{\bibfnamefont{P.}~\bibnamefont{Brax}},
  \bibinfo{author}{\bibfnamefont{C.}~\bibnamefont{van~de Bruck}},
  \bibinfo{author}{\bibfnamefont{A.-C.} \bibnamefont{Davis}},
  \bibinfo{author}{\bibfnamefont{J.}~\bibnamefont{Khoury}}, \bibnamefont{and}
  \bibinfo{author}{\bibfnamefont{A.}~\bibnamefont{Weltman}},
  \bibinfo{journal}{Phys. Rev. D} \textbf{\bibinfo{volume}{70}},
  \bibinfo{pages}{123518} (\bibinfo{year}{2004}), \eprint{astro-ph/0408415}.

\bibitem[{\citenamefont{Agrawal et~al.}(2019)\citenamefont{Agrawal, Cyr-Racine,
  Pinner, and Randall}}]{Agrawal:2019lmo}
\bibinfo{author}{\bibfnamefont{P.}~\bibnamefont{Agrawal}},
  \bibinfo{author}{\bibfnamefont{F.-Y.} \bibnamefont{Cyr-Racine}},
  \bibinfo{author}{\bibfnamefont{D.}~\bibnamefont{Pinner}}, \bibnamefont{and}
  \bibinfo{author}{\bibfnamefont{L.}~\bibnamefont{Randall}}
  (\bibinfo{year}{2019}), \eprint{1904.01016}.

\bibitem[{\citenamefont{Damour and Polyakov}(1994)}]{Damour:1994zq}
\bibinfo{author}{\bibfnamefont{T.}~\bibnamefont{Damour}} \bibnamefont{and}
  \bibinfo{author}{\bibfnamefont{A.~M.} \bibnamefont{Polyakov}},
  \bibinfo{journal}{Nucl. Phys. B} \textbf{\bibinfo{volume}{423}},
  \bibinfo{pages}{532} (\bibinfo{year}{1994}), \eprint{hep-th/9401069}.

\bibitem[{\citenamefont{Gasperini et~al.}(2002)\citenamefont{Gasperini, Piazza,
  and Veneziano}}]{Gasperini:2001pc}
\bibinfo{author}{\bibfnamefont{M.}~\bibnamefont{Gasperini}},
  \bibinfo{author}{\bibfnamefont{F.}~\bibnamefont{Piazza}}, \bibnamefont{and}
  \bibinfo{author}{\bibfnamefont{G.}~\bibnamefont{Veneziano}},
  \bibinfo{journal}{Phys. Rev. D} \textbf{\bibinfo{volume}{65}},
  \bibinfo{pages}{023508} (\bibinfo{year}{2002}), \eprint{gr-qc/0108016}.

\bibitem[{\citenamefont{Brax et~al.}(2010)\citenamefont{Brax, van~de Bruck,
  Davis, and Shaw}}]{Brax:2010gi}
\bibinfo{author}{\bibfnamefont{P.}~\bibnamefont{Brax}},
  \bibinfo{author}{\bibfnamefont{C.}~\bibnamefont{van~de Bruck}},
  \bibinfo{author}{\bibfnamefont{A.-C.} \bibnamefont{Davis}}, \bibnamefont{and}
  \bibinfo{author}{\bibfnamefont{D.}~\bibnamefont{Shaw}},
  \bibinfo{journal}{Phys. Rev. D} \textbf{\bibinfo{volume}{82}},
  \bibinfo{pages}{063519} (\bibinfo{year}{2010}), \eprint{1005.3735}.

\bibitem[{\citenamefont{Oldengott et~al.}(2015)\citenamefont{Oldengott, Rampf,
  and Wong}}]{Oldengott:2014qra}
\bibinfo{author}{\bibfnamefont{I.~M.} \bibnamefont{Oldengott}},
  \bibinfo{author}{\bibfnamefont{C.}~\bibnamefont{Rampf}}, \bibnamefont{and}
  \bibinfo{author}{\bibfnamefont{Y.~Y.~Y.} \bibnamefont{Wong}},
  \bibinfo{journal}{JCAP} \textbf{\bibinfo{volume}{04}}, \bibinfo{pages}{016}
  (\bibinfo{year}{2015}), \eprint{1409.1577}.

\bibitem[{\citenamefont{Khoury and
  Weltman}(2004{\natexlab{a}})}]{Khoury:2003aq}
\bibinfo{author}{\bibfnamefont{J.}~\bibnamefont{Khoury}} \bibnamefont{and}
  \bibinfo{author}{\bibfnamefont{A.}~\bibnamefont{Weltman}},
  \bibinfo{journal}{Phys. Rev. Lett.} \textbf{\bibinfo{volume}{93}},
  \bibinfo{pages}{171104} (\bibinfo{year}{2004}{\natexlab{a}}),
  \eprint{astro-ph/0309300}.

\bibitem[{\citenamefont{Khoury and
  Weltman}(2004{\natexlab{b}})}]{Khoury:2003rn}
\bibinfo{author}{\bibfnamefont{J.}~\bibnamefont{Khoury}} \bibnamefont{and}
  \bibinfo{author}{\bibfnamefont{A.}~\bibnamefont{Weltman}},
  \bibinfo{journal}{Phys. Rev. D} \textbf{\bibinfo{volume}{69}},
  \bibinfo{pages}{044026} (\bibinfo{year}{2004}{\natexlab{b}}),
  \eprint{astro-ph/0309411}.

\end{thebibliography}


\end{document}